\documentclass[aps,prd,showpacs,nofootinbib,floats,floatfix,preprintnumbers,groupedaddress,twocolumn]{revtex4}

\usepackage{graphicx}  
\usepackage{amsmath}   
\usepackage{amssymb}   
\usepackage{color}

\def\eq#1{{Eq.~(\ref{#1})}}

\begin{document}
\title{Emergence and Expansion of Cosmic Space as due to the Quest for Holographic Equipartition}
\author{T. Padmanabhan}
\email{paddy@iucaa.ernet.in}
\affiliation{IUCAA,
Post Bag 4, Ganeshkhind,\\
Pune University Campus, Pune - 411 007, India}

\date{\today}


  \begin{abstract}
One possible interpretation of the holographic principle is the equality of the number of degrees of freedom in a bulk region of space and the number of degrees of freedom on the boundary surface. It is known that such an equality is maintained on equipotential surfaces in any static spacetime in the form of an equipartition law $\Delta N_{\rm bulk}\equiv[\Delta E/ (1/2)k_B T]=  \Delta N_{\rm sur} $. In the cosmological context, the de Sitter universe obeys the same holographic equipartition. I argue that the difference between the surface degrees of freedom and  the bulk degrees of freedom in a region of space (which has already emerged) drives the accelerated expansion of the universe through a simple equation $\Delta V = \Delta t (N_{\rm sur} - N_{\rm bulk})$ where $V$ is the Hubble volume in Planck units and $t$ is the cosmic time in Planck units. 
This equation reproduces the standard evolution of the universe.
This approach provides a novel paradigm to study the emergence of space and cosmology, and has far reaching implications. 
 \end{abstract}

  \maketitle
 
 
 Recent research supports the point of view that gravitational field equations in a wide class of theories should have the same status as the equations of emergent phenomena like fluid mechanics or elasticity. (For a review, see \cite{tp1}; for a small sample of work in the same spirit, see \cite{others}.) In my previous work on the emergent paradigm, I have  treated \textit{only} the field equations as emergent while assuming the existence of a spacetime manifold, metric, curvature etc. as given structures. The field equations then arise as certain consistency conditions obeyed by the background spacetime.

The question arises as to whether one can think of the ``spacetime itself as an emergent structure'' from some suitably defined pre-geometric variables.  While the phrase in quotes above sounds attractive, it is not of much use unless the idea can be translated into mathematical  expressions consistent with what we know about spacetime. 
There are (at least) two key issues which need to be satisfactorily addressed in this context.

To begin with, 
it appears conceptually very difficult to think of time as being emergent from some pre-geometric variable and yet provide a natural way of describing the evolution of dynamical variables. Secondly, it seems facetious  --- if not downright incorrect --- to argue that space is emergent around  \textit{finite} gravitating systems like the Earth, Sun, Milky Way, etc. Any emergent description of the gravitational fields of such systems must accept space around them as pre-existing, just based on physical evidence. Thus, in the study of \textit{finite} gravitating systems without any special status to a time variable, it is extremely difficult to come up with a conceptually consistent formulation for ``spacetime itself as an emergent structure''.

Interestingly enough, both these difficulties disappear when we consider cosmology! Observations indicate that there is indeed a special choice of time variable available in our universe, viz., the proper time of the geodesic observers to whom CMBR appears homogeneous and isotropic. So, one has some justification in treating time differently from space in (and possibly only in) the cosmic context. Second, the spatial expansion of the universe can certainly be thought of as emergence of space as the cosmic time progresses. Thus, in the case of cosmology, we have a well defined setting to make concrete the idea that \textit{cosmic space is emergent as cosmic time progresses}.  

To understand \textit{why} cosmic space emerges --- or, equivalently, to understand why the universe is expanding --- we can use a specific version of holographic principle. To motivate this, let us begin by considering a pure de Sitter universe with a Hubble constant $H$. It is known that such a universe obeys the holographic principle in the form
\begin{equation}
 N_{\rm sur} = N_{\rm bulk}
\label{key1}
\end{equation} 
Here the $N_{\rm sur}$ is the number of degrees of freedom on the spherical surface of Hubble radius  $H^{-1}$ given by
\begin{equation}
 N_{\rm sur} = \frac{4\pi}{L_P^2 H^2}
\end{equation} 
if we attribute one degree of freedom per Planck area.
The $N_{\rm bulk} = |E|/[(1/2)k_BT]$ is the effective number of degrees of freedom which are in equipartition  at the horizon temperature $T=(H/2\pi)$ if we take $|E|$ to be the Komar energy $|(\rho +3p)| V$ contained inside the Hubble volume $V=(4\pi/3H^3)$. That is, 
\begin{equation}
 N_{\rm bulk}=-\frac{E}{(1/2) k_BT} = - \frac{2(\rho +3p)V}{k_BT} 
\label{Nbulk}
\end{equation} 
If we substitute $p=-\rho$, then \eq{key1} reduces to the standard result $H^2= 8\pi L_P^2 \rho/3$, showing the consistency.
The $(\rho +3p)$ is the proper Komar energy density and $V = 4\pi/3H^3$ is the proper volume of the Hubble sphere. The corresponding co-moving expressions will differ by $a^3$ factors in both, which will cancel out.

The condition in \eq{key1} can also be expressed in the form $|E| = (1/2)N_{\rm sur} k_BT$ which is the standard equipartition law that I have obtained  for static spacetimes in GR in 2004 (see, Ref.~\cite{cqg04}) and later generalized to a much wider class of models \cite{tpsurface}. It is therefore appropriate to call the condition \eq{key1} as \textit{holographic equipartition} because it relates the effective degrees of freedom residing in the bulk, determined by the equipartition condition, to the degrees of freedom on the boundary surface.

Our universe, of course, is not exactly de Sitter but there is 
considerable evidence that it is asymptotically de Sitter. This would suggest that the expansion of the universe --- which I consider as conceptually equivalent to the emergence of space --- is being driven towards holographic equipartition. Then the basic law governing the emergence of space must relate the emergence of space (in the form of availability of greater and greater volumes of space) to the difference $(N_{\rm sur} - N_{\rm bulk})$. The most natural and simplest form of such a law will be 
\begin{equation}
 \Delta V = \Delta t (N_{\rm sur} - N_{\rm bulk})
\label{key2}
\end{equation} 
 where $V$ is the Hubble volume in Planck units and $t$ is the cosmic time in Planck units. More generally, one would have expected $(\Delta V/\Delta t)$ to be some function of $(N_{\rm sur} - N_{\rm bulk})$ which vanishes when the latter does.
We could then imagine \eq{key2} as a Taylor series expansion of this function truncated at the first order. 

I will now elevate this relation to the status of a postulate and show that it is equivalent to the standard Friedmann equation. Reintroducing the Planck scale and setting $(\Delta V/\Delta t ) = dV/dt$, this equation becomes
\begin{equation}
 \frac{dV}{dt} = L_P^2 (N_{\rm sur} - N_{\rm bulk})
\end{equation} 
Substituting $V= (4\pi/3H^3), \ N_{\rm sur} = (4\pi/L_P^2 H^2), \ T=H/2\pi$ and using $N_{\rm bulk}$ in \eq{Nbulk}, this equation simplifies to the relation:
\begin{equation}
 \frac{\ddot a}{a}=- \frac{4\pi L_P^2}{3} (\rho + 3p)
\label{frw}
\end{equation} 
which is the standard dynamical equation for the Friedmann model. Using the energy conservation for matter $d(\rho a^3) = -pda^3$ and the de Sitter boundary condition at late times, one gets back the standard accelerating universe scenario. 

The definition of $N_{\rm bulk}$ given in \eq{Nbulk} makes sense only in the accelerating phase of the universe where $(\rho + 3p) <0$ thereby making $N_{\rm bulk}>0$. For normal matter we would not like to have the negative sign in \eq{Nbulk}. This is easily taken care of by using appropriate signs for the two different cases and writing: 
\begin{equation}
 \frac{dV}{dt} = L_P^2 (N_{\rm sur} -\epsilon N_{\rm bulk}); 
\label{key3}
\end{equation} 
and redefining
\begin{equation}
 N_{\rm bulk} = -\epsilon \frac{2(\rho +3p)V}{k_BT} 
\label{nep}
\end{equation} 
with $\epsilon = +1$ if $(\rho + 3p )<0$ and $\epsilon = -1$ if $(\rho + 3p ) >0$. (One could have, of course, used the opposite convention for $\epsilon$ and omitted the minus sign in \eq{nep}; I prefer to maintain the form of \eq{key2} for the accelerating phase of the universe.)
Since only the combination $+\epsilon^2 (\rho+3p)\equiv (\rho+3p)$ occurs in $(dV/dt)$, the derivation of \eq{frw} remains unaffected and we also  maintain $N_{\rm bulk}>0$ in all situations.
 (See Fig. 1.)

\begin{figure}
\label{fig:dof}
\begin{center}
 \scalebox{0.31}{\input{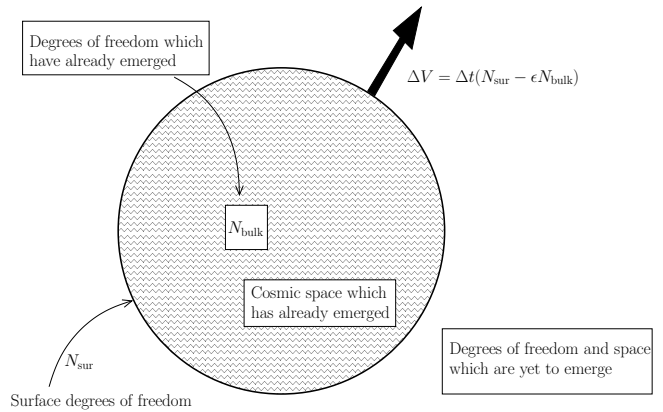}}
\caption{Illustration of the ideas described in this paper. The shaded region is the cosmic space which has already emerged by the cosmic time $t$, along with (a) the surface degrees of freedom ($N_{\rm sur}$) and (b) the bulk degrees of freedom ($N_{\rm bulk}$) that have reached equipartition at the Hubble temperature. The further emergence of cosmic space, measured by the increase in the volume of the Hubble sphere with respect to cosmic time, is driven by the holographic discrepancy ($N_{\rm sur} - \epsilon N_{\rm bulk}$) between these two as indicated by the   equation in the figure. This equation correctly reproduces standard cosmic evolution.}
\end{center}
\end{figure}

We can gain more insights about \eq{key3} by 
separating out the matter component, which causes deceleration, from the dark energy which causes acceleration.
Let us assume, for simplicity, that the universe has two components (matter and dark energy) with $(\rho + 3p)>0$ for matter and $(\rho + 3p)<0$ for  dark energy. Then, \eq{key3} can be written in an equivalent form as
\begin{equation}
 \frac{dV}{dt} = L_P^2 (N_{\rm sur} + N_m - N_{\rm de})
\label{key4}
\end{equation} 
with each of  $N_{\rm sur}, N_m, N_{\rm de}$ being positive (as they should be) with $(N_m - N_{\rm de})=(2V/k_BT)(\rho+3p)_{tot}$.
 We see that if we want $dV/dt \to 0$ asymptotically, we need a component in the universe  with $(\rho + 3p )<0$. A universe without a dark energy component  has no hope of reaching holographic equipartition.  It is also important that the equation of state for the dark energy component is not too  different from that of a cosmological constant so that the universe does not go into a super exponential expansion. 

Stated in this manner, \textit{the existence of a cosmological constant in the universe is required for asymptotic holographic equipartition.}  While these arguments, of course, cannot determine the value of the cosmological constant, the demand of holographic equipartition makes a strong case for its existence. As far as I know, this is more than what any other model has achieved.

Writing the basic equation in the form of \eq{key4}, suggests that there must be an intimate relationship between the matter degrees of freedom and dark energy degrees of freedom. In any fundamental theory of quantum gravity, we expect matter and gravitational degrees of freedom to emerge together and hence such a relationship is expected. 
Discovering this relationship will be a primary challenge for any pre-geometric model since it might even allow us to determine the value of the cosmological constant.

The most obvious and important extension of this work will be to study the growth of perturbations in the universe using a perturbed version of \eq{key4}. To do this formally in a gauge invariant manner, one will need to express the full gravitational field equation as an evolution equation towards holographic equipartition. (I hope to discuss this in detail in a future publication.) But it is fairly straightforward to work out perturbation theory in a specific gauge even without the full formalism. Given the fact that \eq{key4} is identical to \eq{frw}, we do not expect any  surprises in the perturbation theory except for the following feature.

These ideas suggest that we should think of the universe at any given time as being endowed with temperature $T= H/2\pi$ as far as the emergence of space is concerned. This is the temperature related to the degrees of freedom which have already become emergent in the cosmos (from the pre-geometric variables) and should not be confused with the normal kinetic temperature of matter. But since the mathematics is identical, we will expect thermal fluctuations at the temperature $T=H/2\pi$ to be imprinted on any sub-system which has achieved equipartition. This effect will lead to some corrections to the cosmological perturbation theory in the late universe when we do a thermal averaging in a canonical ensemble. This is somewhat similar in spirit to the thermal fluctuations at the de Sitter temperature leaving their imprint on the density fluctuations which were generated during inflation.

To avoid possible misunderstanding, I stress that this is \textit{in addition to (and not instead of)} any imprint of the current Hubble constant $H_0$ on the sub-horizon  cosmic structures  due to standard structure formation processes. 
There are straightforward dynamical effects in various aspects of structure formation (like formation of dark matter halos, cooling of gas, formation of galaxies with flat rotation curves arising due to interplay of baryonic and DM components, etc.) all of which  depend on the background expansion and thus, on $H_0$.  For example, the standard cosmic structure formation scenario does suggest that  (see e.g., \cite{turnerdlb})  a preferred acceleration scale $a_0 =  c H_0$ is imprinted on galactic scale structures. (This is sometimes presented as evidence for MOND, etc. which I believe is unfounded.) 
One can take any standard result in structure formation which depends on $H_0$, and rewrite it in terms of the horizon temperature using $H=2\pi T$, and try to present it in an emergent/thermodynamic language. This is unnecessary and does not add to our insight. It is therefore important to distinguish between (a) rewriting  standard results in terms of the horizon temperature, and (b) deriving genuine effects which arise due to the emergence of cosmic space and the evolution of degrees of freedom towards holographic equipartition. I hope to return to this issue in a future publication.

There are three more comments I want to make regarding \eq{key2}.
First, the simplicity of \eq{key2} is quite striking and it is remarkable that the standard Friedmann equation can be reinterpreted in this form as an evolution towards holographic equipartition. \textit{If the underlying ideas are not correct, it is very difficult to understand why \eq{key2} holds in our universe. }
For example, consider the definition of horizon temperature in a non-de Sitter universe. In obtaining $N_{\rm bulk}$  we have assumed that the relevant temperature is given by $T= H/2\pi$. Defining a horizon temperature when $H$ is time-dependent  is not easy and there is some amount of controversy in the literature regarding the correct choice. It is possible to obtain equations similar to \eq{key2} with other definitions (all of which agree in the de Sitter limit) but none of them seems to have the compelling naturalness which \eq{key2} possesses. The same is true as regards the volume element $V$ which I took as the Hubble volume; other choices are possible but then the resulting equation has no simple interpretation.

Second, this equation presents the emergence of cosmic space as an evolution towards holographic equipartition in a ``bit-by-bit'' increase in Planck units. When the cosmic time changes by one Planck unit, the amount of extra cosmic space which becomes available changes by $(N_{\rm sur} - \epsilon N_{\rm bulk})$. This is reminiscent of ideas in which one considers cosmic expansion as a computational process. Presented in Planck units, \eq{key2} has no adjustable parameters and opens the possibility for interpretation in combinatorial terms when we understand the pre-geometric variables better.

Third, we expect that when the relevant degrees of freedom like $N_{\rm sur}$ are of the order of unity, \eq{key2} must break down. Studying possible modifications of this equation when the degrees of freedom are few will help us to study the evolution of the universe close to the big bang in a quantum cosmological setting --- but without the usual complications related to the time coordinate. In other words, postulating well motivated corrections to the ``bit dynamics'' expressed by \eq{key2} may be a better way of tackling the singularity problem, rather than by quantizing gravity in the usual manner.

To conclude, let me summarize how these ideas fit in a broader context. 

The degrees of freedom are the basic entities in physics and the holographic principle suggests a deep relationship between the number of degrees of freedom residing in a bulk region of space and the number of degrees of freedom on the boundary of this region. Given a fundamental area  scale, $L_P^2$, it makes sense to count the surface degrees of freedom as $A/L_P^2$ where $A$ is the area of the surface.
The non-trivial task is to come up with a suitable measure for the bulk degrees of freedom which must depend on the matter residing in the bulk. (This necessary dependence on the matter variables precludes counting the bulk degrees of freedom as $V/L_P^3$.) 
It is here that the idea of equipartition comes to our aid. If the surface is endowed with a horizon temperature $T$, then we can think of $E/(1/2)k_BT$ as the number of \textit{effective} bulk degrees of freedom. The logic behind this identification is as follows. We assume that the bulk region is like the inside of a microwave oven with the temperature set to the surface value, as far as the degrees of freedom which have \textit{already emerged} (along with the space) are concerned. Then, if \textit{these} degrees of freedom account for an energy $E$, it follows that  $E/(1/2)k_BT$ is indeed the correct count for effective $N_{\rm bulk}$. This temperature $T$ and $N_{\rm bulk}$ should not be confused with the normal kinetic temperature of matter residing in the bulk and the standard degrees of freedom we associate with matter. It is more appropriate to think of these degrees of freedom as those which have already emerged, along with space, from some pre-geometric variables. 
I postulate  that the emergence of cosmic space is driven by the holographic discrepancy $(N_{\rm sur}  + N_m - N_{\rm de})$ between the surface and bulk degrees of freedom
where $N_m$ is contributed by normal matter with $(\rho + 3p)>0$ and $N_{\rm de}$ is contributed by the cosmological constant with all the degrees of freedom being counted positive.
It is obvious that in the absence of $N_{\rm de}$, this expression can never be zero and holographic equipartition cannot be achieved. In the presence of the cosmological constant, the emergence of space will soon lead to $N_{\rm de}$ dominating over $N_m$ with the universe commencing accelerated expansion. Asymptotically, $N_{\rm de}$ will approach $N_{\rm sur}$ 
and the rate of emergence of space, $dV/dt$, will tend to zero in a de Sitter universe. The cosmos would have found its peace. 

The expansion which we witness (which is equivalent to emergence of space) is an evolution towards equilibrium of a system which is right now away from equilibrium. Such an evolution makes perfect thermodynamic sense. Needless to say, this approach provides a completely different paradigm to study cosmology.


\end{document}